\documentclass[prd,reprint,superscriptaddress]{revtex4-2}
\pdfoutput=1 

\usepackage{hyperref}
\hypersetup{colorlinks=true, citecolor=blue, urlcolor=blue, linkcolor=blue}

\usepackage[T1]{fontenc} 

\usepackage{graphicx}
\usepackage{amssymb}
\usepackage{amsmath}
\usepackage{color}
\usepackage{multirow}
\usepackage{float}
\usepackage{caption}
\usepackage{subcaption}
\DeclareMathAlphabet\mathbfcal{OMS}{cmsy}{b}{n}
\usepackage{cancel}
\usepackage{soul}

\begin{document} 
\title{Dark Photon Bremsstrahlung and Ultra-High-Energy Cosmic Ray}


\author{Predee Tantirangsri}
\email{predeetan@gmail.com}
\affiliation{Department of Physics, Mahidol University, Bangkok 10400, Thailand}
\author{Daris Samart}
\email{darisa@kku.ac.th}
\affiliation{Khon Kaen Particle Physics and Cosmology Theory Group (KKPaCT), Department of Physics, Khon Kaen University, 123 Mitraphap Rd., Khon Kaen, 40002, Thailand}
\affiliation{National Astronomical Research Institute of Thailand, Chiang Mai 50180, Thailand}
\author{Chakrit Pongkitivanichkul}
\email{chakpo@kku.ac.th}
\affiliation{Khon Kaen Particle Physics and Cosmology Theory Group (KKPaCT), Department of Physics, Khon Kaen University, 123 Mitraphap Rd., Khon Kaen, 40002, Thailand}
\affiliation{National Astronomical Research Institute of Thailand, Chiang Mai 50180, Thailand}




\begin{abstract}
A dark photon is a hypothetical particle that is similar to a photon with a small mass and interacts very weakly with ordinary matter through a kinetic mixing with the ordinary photon. In this paper, we propose a new way to probe the existence of dark photons through the Bremsstrahlung effect on ultra-high-energy cosmic rays (UHECRs). Using the standard soft photon calculation, we demonstrate that the dark photon Bremsstrahlung process could lead to significant energy loss for protons in the ultralight dark photon scenario, and that this effect could be tested against observational data of UHECRs. We also provide exclusion limits which can be compared with existing limits on ultralight dark photons.
\end{abstract}

\maketitle


\section{Introduction} \label{sec:intro}
Cosmic rays are charged particles coming from sources beyond our solar system. Since its first discovery a century ago, the sources and propagation mechanisms still remain the active research topics. The energy range of cosmic rays spans over many orders of magnitude ($10^6$ eV to $10^{20}$ eV). This aspect of the cosmic rays has been attractive since it involves the high energy processes which could potentially hint us towards a new physics scale.


The general feature of the cosmic rays spectrum can be described by a broken power law $\frac{dN}{dE} \propto E^{-\alpha}$, where $\alpha$ is a spectral index responsible for a particular range of energies (see \cite{2009PrPNP..63..293B} for extensive review). At the energies below several PeV scale, the spectral index has been measured to be $\alpha \approx 2.75$. Then the steepening of the spectrum is observed at $E \sim 10^{15.5}$ eV which is also known as the ``knee". The spectrum continues with $\alpha \approx 3.1$ up to the ``angle" with $E \sim 4 \times 10^{17}$ eV. After that it becomes less steep with $\alpha \approx 2.2-2.5$ up to the very tail end of the observed spectrum $\sim 10^{20}$ eV. Ultra-high-energy cosmic rays (UHECR) are defined as cosmic rays with energy higher than $10^{19}$ eV. There are many unknown aspects of the UHECR, in particular, the origin of cosmic rays at these energies has been challenging so far. Although some possible sources of the UHECR require no new physics such as neutron stars, active galactic nuclei (AGN) \cite{Mahajan:2013lia,Amenomori:2019rjd}, there have been interesting proposals from beyond standard model theories of particle physics (BSM) such as dark matter, axion and extra dimension \cite{Kazanas:2001mr,Grib:2007np,Nicolaidis:2009zg,Dixit:2010yur}.

Despite the fact that the Large Hadron Collider (LHC) has been running for a decade, there is still no new concrete evidence of any new physics yet. One interesting possibility is that new physics might not interact via usual standard model gauge interactions. The interaction with new physics could come indirectly from the mixing between gauge bosons instead. If the new physics sector has its own $U(1)$ gauge symmetry, the kinetic mixing term with the standard model photon lead is always allowed. Therefore the existence of new $U(1)$ gauge boson known as dark photon \cite{HOLDOM1986196,FAYET1980285,FAYET1990743,Fabbrichesi:2020wbt,Caputo:2021eaa} could be probed via the mixing term with the usual photon. In past decades, many attempts have been made to detect dark photon, such as gravitational wave detection \cite{Guo:2019ker,Co:2021rhi}, helioscope \cite{Redondo:2008aa,Ehret:2010mh,Betz:2013dza}, and spectroscopy \cite{Jaeckel:2010xx,Danilov:2018bks}. Detecting dark photon could also be the signal of the dissipative dark sector \cite{Foot:2014uba,Foot:2014osa,Foot:2016wvj}.

In this paper, we propose another way to probe the dark photon by considering the dark photon Bremsstrahlung process in the UHECR. We show that the energy loss of protons from soft and hard dark photon Bremsstrahlung are significant and hence could change the UHECR spectrum. Using the simple broken power law consisting of two astronomical sources as the original flux of cosmic rays, the simple model of dark photon Bremsstrahlung can be tested against the observational data of UHECR. We also provide the exclusion limits from our study and compare them with existing limits on ultralight dark photon.

This paper is organised as follows: In section \ref{sec-2}, the mixing dark photon and photon is presented and we provide all analytical expressions of the soft dark photon Bremsstrahlung from proton scattering. The free parameters of our model will be fitted with the relevant experimental data from UHECR and the results are discussed in section \ref{sec-3}. In section \ref{sec-4}, we close the paper with the conclusion of our findings in this work.





\section{Dark Photon Radiation From Proton Scattering}
\label{sec-2}

Since $90 \%$ of cosmic rays composition are protons, we use the dark photon bremsstrahlung from protons as the main source of dark photon radiation. Let us review the calculation of an ordinary photon bremsstrahlung in this section.

\subsection{Soft photon Bremsstrahlung}\label{sub:softphoton}

\begin{figure}[h!]
    \centering
    \includegraphics[width=0.4\textwidth]{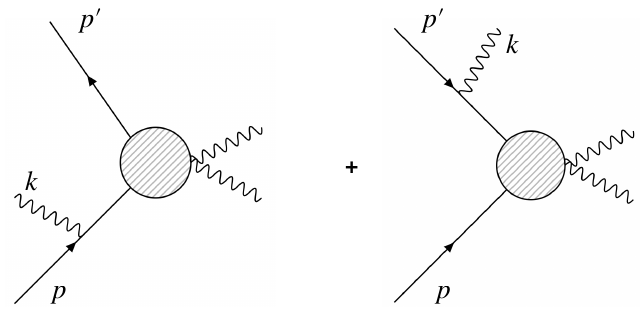}
     \caption{The diagram of soft photon radiated from the proton scattering.}
     \label{fig:pscattering}
 \end{figure}

The soft bremsstrahlung is the process of photons radiating from charged particles \cite{Weinberg:1970bs,Peskin:1995ev}. The main process we are considering is the proton scattering which could induce the soft photon radiation from initial/final states shown in Fig.(\ref{fig:pscattering}). Note that the detail of a hard process involved in the diagram is not relevant to the calculation. In the cosmic ray scattering, these hard processes include inverse Compton scattering, proton-proton or proton-nucleon scattering.
Let's consider first the amplitude from the diagrams
\begin{eqnarray}
    i\mathcal{M} &=& -ie\Bar{u}(p')\left( \mathcal{M}_0(p',p-k)\frac{i(\cancel{p}+\cancel{k}+m)}{(p-k)^2-m^2}\gamma^{\mu}\epsilon_{\mu}^*(k) \right. \nonumber \\
    & & \left. + \gamma^{\mu}\epsilon_{\mu}^*(k)\frac{i(\cancel{p'}+\cancel{k}+m)}{(p'+k)^2 - m^2}\mathcal{M}_0(p'+k,p) \right)u(p),
\end{eqnarray}
where $\mathcal{M}_0$ is the amplitude for proton scattering. Since this photon is soft in the sense that the momentum is much smaller than those of protons, we can approximate
\begin{align}
    |\mathbf{k}| \ll& |\mathbf{p'} - \mathbf{p}|\\
    \mathcal{M}_0(p',p-k) \approx& \mathcal{M}_0(p'+k,p) \approx \mathcal{M}_0(p',p). \nonumber
\end{align}
After the approximation, we can obtain the amplitude for our processes
\begin{align}
    i\mathcal{M} = \Bar{u}(p')[\mathcal{M}_0(p',p)]u(p) \cdot \left[ e\left( \frac{p' \cdot \epsilon^*}{p' \cdot k} - \frac{p \cdot \epsilon^*}{p \cdot k} \right) \right].
\end{align}
Next the phase-space integration is performed on the photon momentum and photon polarizations. The cross section becomes
\begin{align}
    d\sigma(p \rightarrow p' + \gamma) &= d\sigma(p\rightarrow p') \cdot \int \frac{d^3k}{(2 \pi)^3} \frac{1}{2k} \times \nonumber \\
    & \sum_{\lambda=1,2}e^2 \left| \frac{p'\cdot \epsilon^{(\lambda)}}{p'\cdot k} - \frac{p\cdot \epsilon^{(\lambda)}}{p \cdot k} \right|^2,
\end{align}
where $\epsilon^{\lambda}$ is the transverse unit polarization vectors. Therefore, the differential probability of a single photon radiation is then
\begin{align}
    d\sigma = \frac{d^3k}{(2 \Pi)^3} \sum_{\lambda} \frac{e^2}{2k} \left| \boldsymbol{\epsilon}_{\lambda} \cdot \left( \frac{\mathbf{p'}}{p' \cdot k} - \frac{\mathbf{p}}{p\cdot k} \right) \right|^2.
\end{align}
Next, we sum over the polarization vectors and replace it with the identity: $ \sum \epsilon_{\mu} \epsilon_{\nu} = - g_{\mu\mu}$. The probability becomes
\begin{align}
    d\sigma = \frac{d^3k}{(2\pi)^3} \frac{e^2}{2k} \left( \frac{2p \cdot p' }{(k \cdot p')(k \cdot p)} - \frac{m^2}{(k \cdot p')^2} - \frac{m^2}{(k \cdot p)^2} \right). \nonumber
\end{align}
Choosing a frame in which $p^0 = p'^0 = E$, we obtain the set of 4-momenta as
\begin{align}
    k^{\mu} &= (k, \mathbf{k})\nonumber \\
    p^{\mu} &= E(1, \mathbf{v})\\
    p'^{\mu} &= E(1, \mathbf{v'}). \nonumber
\end{align}

Although photon receives an effective mass via interaction with plasma \cite{Caputo:2020bdy,Caputo:2020rnx,Mirizzi:2009iz,Kunze:2015noa},
the above assumption on photon momentum still holds since the energy scale involved in the cosmic ray processes is much larger than the effective photon mass. This assumption is also valid in the case of massive dark photon.
Then the total probability becomes 
\begin{align}
    \sigma = \frac{\alpha}{\pi} \int_0^{|\mathbf{q}|} dk \frac{1}{k} \mathcal{I}(\mathbf{v},\mathbf{v'}), 
\end{align}
where $|\mathbf{q}|=|\mathbf{p}-\mathbf{p'}.|$ is the maximum 3 momentum for the photon.
The function $\mathcal{I}$ is the differential intensity of the photon, $dE_{rad}/dk$, defined as
\begin{align}
    \mathcal{I}(\mathbf{v},\mathbf{v'}) &= \int \frac{d \Omega_{\hat{k}}}{4\pi} \left( \frac{2(1 - \mathbf{v} \cdot \mathbf{v'})}{(1 - \hat{k} \cdot \mathbf{v})(1 - \hat{k} \cdot \mathbf{v'})} \right. \nonumber \\
    & \left. - \frac{m^2 / E^2 }{(1 - \hat{k} \cdot \mathbf{v'})^2} - \frac{m^2/E^2}{(1 - \hat{k} \cdot \mathbf{v})^2}  \right),
\end{align}
which is independent of $k$. We now can break the integral into two parts for two peaks of radiated energies $\hat{k} \cdot \mathbf{v}$ and $\hat{k} \cdot \mathbf{v'}$. Then let $\theta = 0$ along each peak and perform the integrals over $\theta = 0$, we have 
\begin{align}
    \mathcal{I}(\mathbf{v},\mathbf{v}') \sim& \int_{\hat{k} \cdot \mathbf{v}=\mathbf{v}' \cdot \mathbf{v}}^{\cos \theta=1} d\cos \theta \frac{1-\mathbf{v} \cdot \mathbf{v}'}{(1-v\cos \theta)(1 -\mathbf{v} \cdot \mathbf{v}')} \\
   +& \int_{\hat{k} \cdot \mathbf{v}'=\mathbf{v}' \cdot \mathbf{v}}^{\cos \theta = 1} d \cos \theta \frac{1 - \mathbf{v} \cdot \mathbf{v}'}{(1 - \mathbf{v} \cdot \mathbf{v}')(1 - v' \cos \theta)}. \nonumber
\end{align}
After performing the integrals, we get
\begin{align}
    \mathcal{I}(\mathbf{v},\mathbf{v}') &\approx \log \left( \frac{1-\mathbf{v} \cdot \mathbf{v'}}{1 - |\mathbf{v}|} \right) + \log \left(\frac{1-\mathbf{v} \cdot \mathbf{v'}}{1 - |\mathbf{v'}|}\right) \nonumber \\
    &= 2\log \left( \frac{-q^2}{m^2} \right),
\end{align}
where $q^2 = (p'-p)^2$.
Thus, the differential cross section of photon radiation is written as the Sudakov double logarithm:
\begin{equation}
    d\sigma_{p \rightarrow p'+\gamma} \approx d\sigma_{p \rightarrow p'} \cdot \frac{\alpha}{\pi} \ln\left(\frac{-q^2}{\mu^2}\right) \ln\left(\frac{-q^2}{m_p^2}\right)
\end{equation}
where $m_p$ is the proton mass. However, this is not the only contribution to the physical scattering process of $p' \rightarrow p$ in the infrared (IR) limit. There is a loop contribution to the proton vertex in which $\mu$ is treated as the IR cutoff scale of the process. In the limit $\mu \rightarrow 0$, the cross section to the first order of $\alpha$ is written as
\begin{equation}
    d\sigma_{p'\rightarrow p} \approx d\sigma_0 \left[1-\frac{\alpha}{\pi} \ln\left(\frac{-q^2}{\mu^2}\right) \ln\left(\frac{-q^2}{m_p^2}\right)\right].
\end{equation}
Although the soft photon radiation and the IR correction are both divergent, the observed cross section coming from the combined contribution is finite. For any real experiment, a measurement can only detect photons above a certain energy limit, $E_l$. This results in the finite observed cross section
\begin{align}
    d\sigma_{\text{measure}} &\approx d\sigma_0 \left[1-\frac{\alpha}{\pi} \ln\left(\frac{-q^2}{\mu^2}\right) \ln\left(\frac{-q^2}{m_p^2}\right) \right.\nonumber \\
    & \quad\quad + \left. \frac{\alpha}{\pi} \ln\left(\frac{E_l^2}{\mu^2}\right) \ln\left(\frac{-q^2}{m_p^2}\right)\right].\nonumber\\
    &\approx d\sigma_0\left[1-\frac{\alpha}{\pi} \ln\left(\frac{-q^2}{E_l^2}\right) \ln\left(\frac{-q^2}{m_p^2}\right)\right].
\end{align}
However, this result comes from only the leading order correction. It turns out that if we consider to the correction to all order the result becomes
\begin{equation}\label{eq:1}
     d\sigma = d\sigma_0 \times \text{exp}\left[-\frac{\alpha}{\pi} \ln\left(\frac{-q^2}{E_l^2}\right) \ln\left(\frac{-q^2}{m_p^2}\right)\right].
\end{equation}


\subsection{Dark Photon Bremsstrahlung}   
Next we will set up a simple model of dark photon. We start from the Lagrangian of photon and dark photon sector respecting $U(1)_{\rm EM}$ and $U'(1)$ gauge symmetries
\begin{equation} \label{eq:2}
    \mathcal{L} = -\frac{1}{4}F_{\mu \nu}F^{\mu \nu} - \frac{1}4\widetilde{F}_{\mu \nu}\widetilde{F}^{\mu \nu} - \frac{\epsilon}{2}F_{\mu \nu}\widetilde{F}^{\mu \nu}.
\end{equation}
The last term of the Lagrangian is a mixing term between 2 gauge bosons allowed by $U(1)$ and $U'(1)$. 
\begin{figure}[h!]
     \centering
     \includegraphics[width=0.2\textwidth]{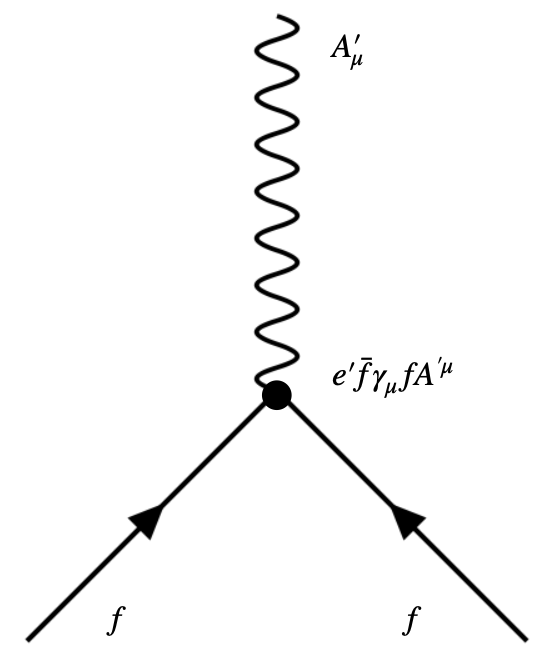}
     \caption{Diagram for dark photon coupling with charge particles}
     \label{fig:darkphotoncoupling}
\end{figure} 
For the massive dark photon case, gauge bosons are assumed to acquire masses via the Stueckelberg Lagrangian,
\begin{equation}
    \mathcal{L} = -\frac{1}{2}M_a^2 A_{a\mu} A^{\mu}_a -\frac{1}{2}M_b^2 A_{b\mu} A^{\mu}_b - M_a M_b A_{a\mu} A^{\mu}_b.
\end{equation}
After the diagonalization of the kinetic terms, we can write the dark photon Lagrangian as
\begin{align}
    \mathcal{L} &= - \frac{1}4F'_{\mu \nu}F'^{\mu \nu} + \frac{1}{2}m_{\gamma'}^2A'_{\mu}A'^{\mu} \\
    & - \frac{1}{\sqrt{1-2\delta\epsilon+\delta^2}} \frac{e (\delta -\epsilon)}{\sqrt{1-\epsilon^2}} A'^{\mu} J^{\rm SM}_{\mu}, \nonumber
\end{align}
where $A'_{\mu}$ is the dark gauge boson field, $e$ is the usual charge in the normal EM theory, $\delta = \frac{M_a}{M_b}$ and $m_{\gamma'}^2 = M_a^2 + M_b^2$. 
Note that the presence of the mass terms removes the freedom of choosing how dark photon is coupled to the standard model current even when the mass matrix is already diagonal, i.e., $M_b \rightarrow 0$. For our effective approach, we simply assume that the Stueckelberg mechanism gives mass to only one of the $U(1)$ gauge boson ($\delta = 0$) and the effective coupling between the dark photon and the standard model charged current $J^{\rm SM}_{\mu}$ becomes
\begin{align}
    &\mathcal{L} = -e'A'^{\mu} J^{\rm SM}_{\mu},\nonumber\\
    &e' \equiv - \frac{e \epsilon}{\sqrt{1-\epsilon^2}},  \quad \alpha' \equiv \frac{e'^2}{4\pi} = \frac{\alpha\epsilon^2}{1-\epsilon^2}, \label{eq:alphamix}
\end{align}
where $\alpha = e^2/4\pi = 1/137$ is the standard EM fine structure constant. Note that in the vanishing mass limit of dark photon, the dark photon still maintains a non-zero coupling with the standard model current as discussed in \cite{Fabbrichesi:2020wbt}. Due to this reason, this type of Lagrangian is mostly used in the experimental limits of dark photon.



The extra contribution from dark photon radiation can be treated as the energy loss function as following.
Assume that the energy of the incoming proton is $\sqrt{s_0}$. After dark photon radiation, the energy of the outgoing proton becomes $\sqrt{s}$. Therefore the 4 momenta of the incoming proton and outgoing proton are written as
\begin{equation}
    p^{\mu} = \left(\frac{\sqrt{s_0}}{2},\mathbf{p}\right),\quad p'^{\mu} = \left(\frac{\sqrt{s}}{2},\mathbf{p'}\right).
\end{equation}
Note that $s$ and $s_0$ are not necessary the center of mass energies of the scattering process.

The loss function is defined as the multiplication factor in the cross section as
\begin{equation}
    d\sigma(p \rightarrow p + \gamma')  = d\sigma_0 F(s,y),
\end{equation}
where $d\sigma_0$ is the scattering process of the protons. The energy of dark photon $E$ and the fraction $y$ are defined as
\begin{equation}\label{eq:6}
    E \equiv \frac{\sqrt{s}-\sqrt{s_0}}{2},\; y \equiv \frac{E}{\frac{\sqrt{s}}{2}}.
\end{equation}

\begin{figure}[ht]
     \centering
     \includegraphics[width=0.15\textwidth]{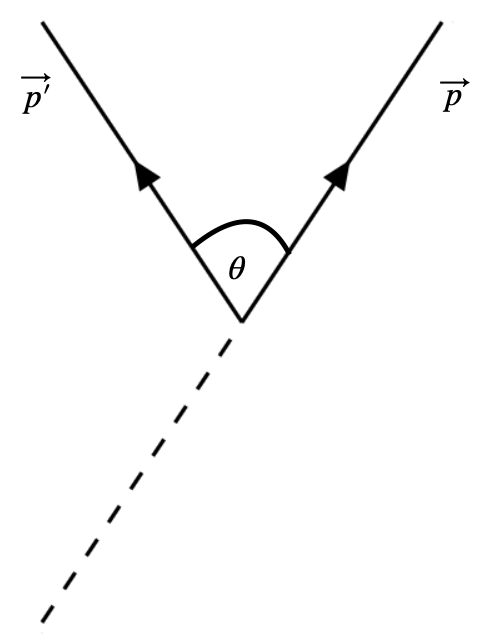}
     \caption{The angle $\theta$ is defined as an angle between incoming momentum ($\mathbf{p}$) and outgoing momentum ($\mathbf{p'}$)}
     \label{fig:theta}
 \end{figure}

In the case of soft dark photon bremsstrahlung from proton scattering, the energy of dark photon is in the range of low energy, i.e., $m_{\gamma'} < E < m_p$ where the lowest possible energy of the dark photon is its own mass and the highest energy considered soft is the mass of the proton.
In order to find the loss function for soft dark photon, we rewrite Eq.(\ref{eq:1}) by replacing the lowest energy threshold with the dark photon mass, $E_l \rightarrow m_{\gamma'}$ and the EM fine structure constant with the dark photon one. The cross section becomes 
\begin{equation}\label{eq:3}
     d\sigma = d\sigma_0 \times \text{exp}\left[-\frac{\alpha'}{\pi} \ln\left(\frac{-q^2}{m_{\gamma'}^2}\right) \ln\left(\frac{-q^2}{m_p^2}\right)\right].
\end{equation}
Notice that the effect of soft dark photon radiation is absent when the mass of the dark photon vanishes ($d\sigma \rightarrow 0$ as $m_{\gamma'}\rightarrow 0$). This is the dark photon decoupling limit for the soft radiation since the massless dark photon IR divergent is perfectly cancelled by the soft radiation leaving no observable effect. 
The transfer momentum can then be written as
\begin{align}
    -q^2 &= -(p_{\mu}'-p_{\mu})^2 \\
        &= \frac{\sqrt{s s_0}}{2} - \frac{1}{2}\sqrt{s_0 - 4m_p^2}\sqrt{s-4m_p^2}\cos{\theta} - 2m_p^2 \nonumber
\end{align}
where $\theta$ is the angle between $p$ and $p'$ as shown in Fig.(\ref{fig:theta}).

Then integrate all over possible values of the scattering angle, the loss function becomes

\begin{align}
   &F(s,y) = \int^{\pi}_0 d\theta\;\;\text{exp}\left[-\frac{\alpha'}{\pi} \times \right. \label{eq:softLF} \\
   &\ln\left(\frac{\frac{s(1-y)}{2} - \frac{1}{2}\sqrt{s(1-y)^2 - 4m_p^2}\sqrt{s-4m_p^2}\cos{\theta} - 2m_p^2}{m^2_{\gamma'}}\right) \nonumber \\
    & \left.\ln\left(\frac{\frac{s(1-y)}{2} - \frac{1}{2}\sqrt{s(1-y)^2 - 4m_p^2}\sqrt{s-4m_p^2}\cos{\theta} - 2m_p^2}{m^2_{p}}\right) \right]. \nonumber
\end{align}

For the hard dark photon Bremsstrahlung case, the energy of the dark photon is higher than the assumption made in \ref{sub:softphoton}. In the energy range $E > m_p$, we use the following relation instead:
\begin{equation}
    d\sigma(p \rightarrow p + \gamma') = d\sigma_0 \times \frac{\alpha'}{\alpha}.
\end{equation}
The reason behind this assumption is that the cross section of the dark photon radiation should be similar to the one with the photon radiation in the high energy limit except for the different coupling. Therefore the simple scaling is assumed in the losing function for the hard dark photon Bremsstrahlung,
\begin{equation}
    F(s,y) = \frac{\alpha'}{\alpha}. \label{eq:hardLF}
\end{equation}

\begin{figure}
    \centering
    \includegraphics[width=\linewidth]{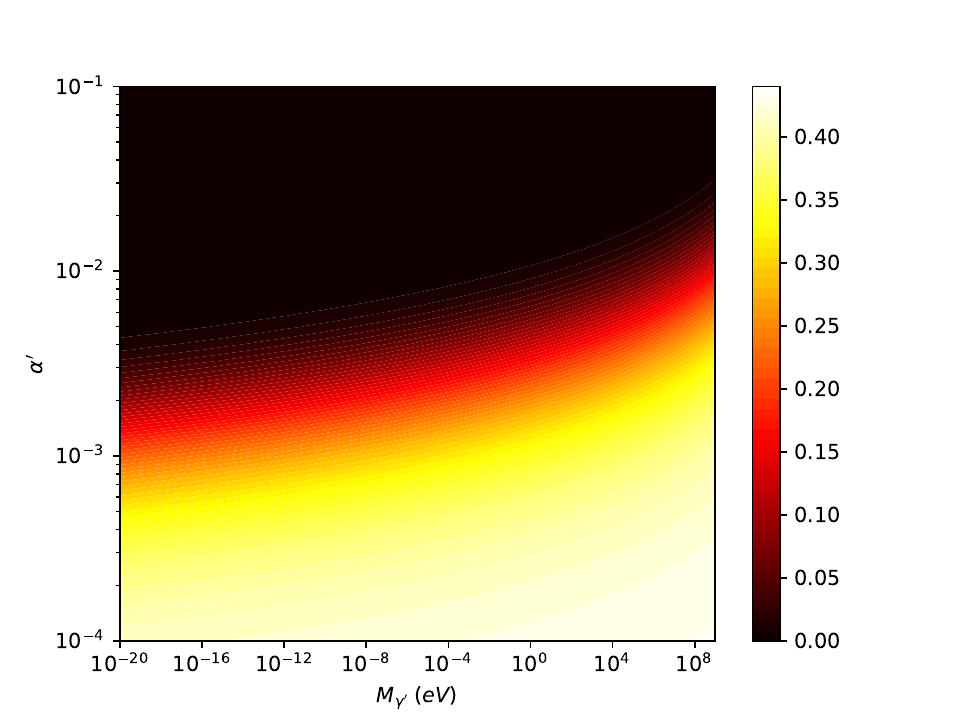}
    \includegraphics[width=\linewidth]{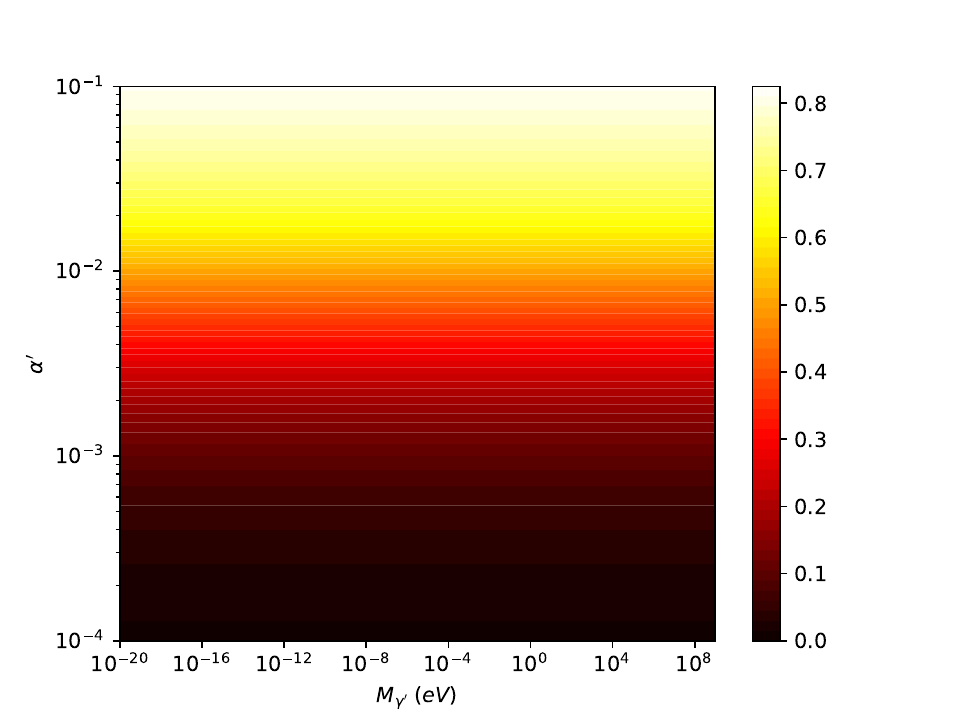}
    \caption{The losing factor as a function of $\alpha'$ and $m_{\gamma'}$. The top panel is for soft dark photon Bremsstrahlung where $s = 10^{19}$ eV and $E = 0.1$ GeV. The bottom panel is for hard dark photon Bremsstrahlung where $s = 10^{19}$ eV and $E = 10$ GeV.}
    \label{fig:LF}
\end{figure}

Now consider the probability of a proton with energy, $\sqrt{s}$, losing a fraction of energy $y$ due to the dark photon radiation which is written as
\begin{equation}
    P(s,y) = \frac{\text{Events with dark photon radiation}}{ \text{All events}}.
\end{equation}
Interpreting this in terms of the cross section, we define the losing factor, $LF(s,y)$, as
\begin{align}\label{eq:7}
    LF(s,y) &\equiv P(s,y) = \frac{\sigma(p \rightarrow p + \gamma') }{\sigma_0 + \sigma(p \rightarrow p + \gamma') } \nonumber \\
    & = \frac{F(s,y)}{1 + F(s,y)}.
\end{align}
In Fig.(\ref{fig:LF}), the losing factor for soft dark photon and hard dark photon have been shown as a contour plot of $\alpha'$ and $m_{\gamma'}$. Note that for the soft dark photon effect vanishes when $m_{\gamma'}\rightarrow 0$ and $\alpha' \rightarrow 1$. However, for the hard dark photon effect which is dominating over the larger range of energy, the losing factor tends to zero as $\alpha' \rightarrow 0$ as expected. Another interesting feature about the hard photon effect is that the losing factor is independent of the dark photon mass. This is because the presence of an almost massless dark photon contribute similarly to photon ones in the proton scattering. In fact if we assume that the coupling $\alpha' = \alpha$ the losing factor is exactly $0.5$ as the photon and dark photon radiation is twice as large as the photon radiation alone.

The direct result of this effect on the flux of protons on a particular energy, $\sqrt{s_0}$, is that there are 2 contributions to the original cosmic rays spectrum, i.e., the increase in flux due to protons from a higher energy radiating dark photon and the decrease in flux due to protons at that energy lose energy radiating dark photon.
Thus, we can calculate the cosmic rays spectrum at energy, $\sqrt{s_0}$, using the factor of losing energy,
\begin{align}\label{eq:8}
    I(s_0) &= \int_{s_0}^{\infty}ds I_0(s)LF(s,y)\frac{dy}{ds} \\
    &+ I_0(s_0)\left( 1 - \int_{(2m_p+m_{\gamma})^2}^{s_0}ds\; LF(s_0,y')\frac{dy'}{ds} \right), \nonumber
\end{align}
where $y = \frac{E}{\frac{\sqrt{s}}{2}} = 1 - \sqrt{\frac{s_0}{s}}$ and $y' = \frac{E}{\frac{\sqrt{s_0}}{2}}= 1 - \sqrt{\frac{s}{s_0}}$.

The Eq.(\ref{eq:8}) is the prediction of cosmic rays spectrum at energy $s_0$.  
The first term shows the increased spectrum at energy $s_0$ from proton radiating dark photon at higher energy at $s > s_0$, and the second term represents the drop in spectrum corresponding to dark photon radiation at $s = s_0$. The original spectrum, $I_0$, is the incident cosmic ray spectrum, which has two components, cosmic rays originated from extragalactic sources ($I_E$) and cosmic rays originated within our galaxy ($I_G$) \cite{Kazanas:2001ep,PierreAuger:2021hun,Ivanov:2020rqn}

\begin{align}
     I_0 &= I_G + I_E \label{eq:10}\\
     I_E &= \mathcal{I}_{E}^{(0)}\left(\sqrt{s_0}\right)^{-\alpha_E} \times \exp{\left(-\sqrt{s_0}/E_E\right)} \nonumber \\
     I_G &= \mathcal{I}_{G}^{(0)}\left(\sqrt{s_0}\right)^{-\alpha_G} \times \exp{\left(-\sqrt{s_0}/E_G\right)}\nonumber,
\end{align}
where $\mathcal{I}_{E}^{(0)}$ and $\mathcal{I}_{G}^{(0)}$ are proportional constants. 
The spectral indices and cut-off energies for Telescope Array \cite{Ivanov:2020rqn} and Pierre Auger Observatory \cite{PierreAuger:2021hun} are given by $(\alpha_E, \alpha_G, E_E, E_G) = (2.68, 3.28, 10^{10.81} \;\text{GeV}, 10^{9.69} \;\text{GeV})$ and $(\alpha_E, \alpha_G, E_E, E_G) = (2.52, 3.30, 4.7 \times 10^{10} \;\text{GeV}, 4.9\times10^{9} \;\text{GeV})$ respectively. Therefore, the parameters in this model, $\{\alpha', m_{\gamma'}, \mathcal{I}_{E}^{(0)}, \mathcal{I}_{G}^{(0)}\}$, will be subjected to the experimental data in the next section.


\section{Results and Discussion}
\label{sec-3}

\begin{figure}[!ht]
     \centering
     \includegraphics[width=0.5\textwidth]{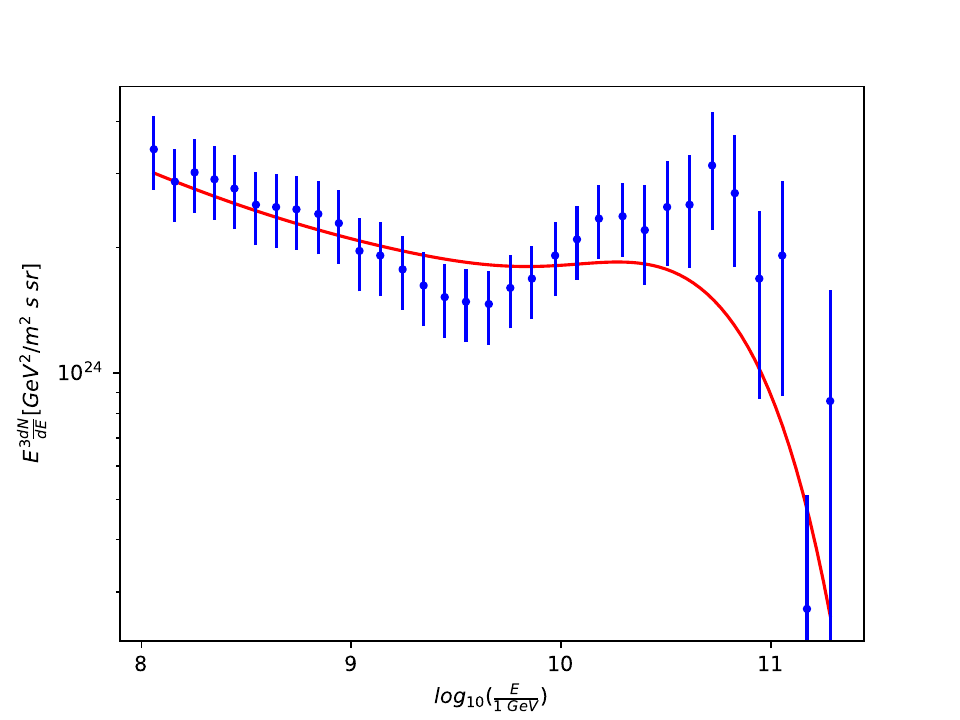}
     \includegraphics[width=0.5\textwidth]{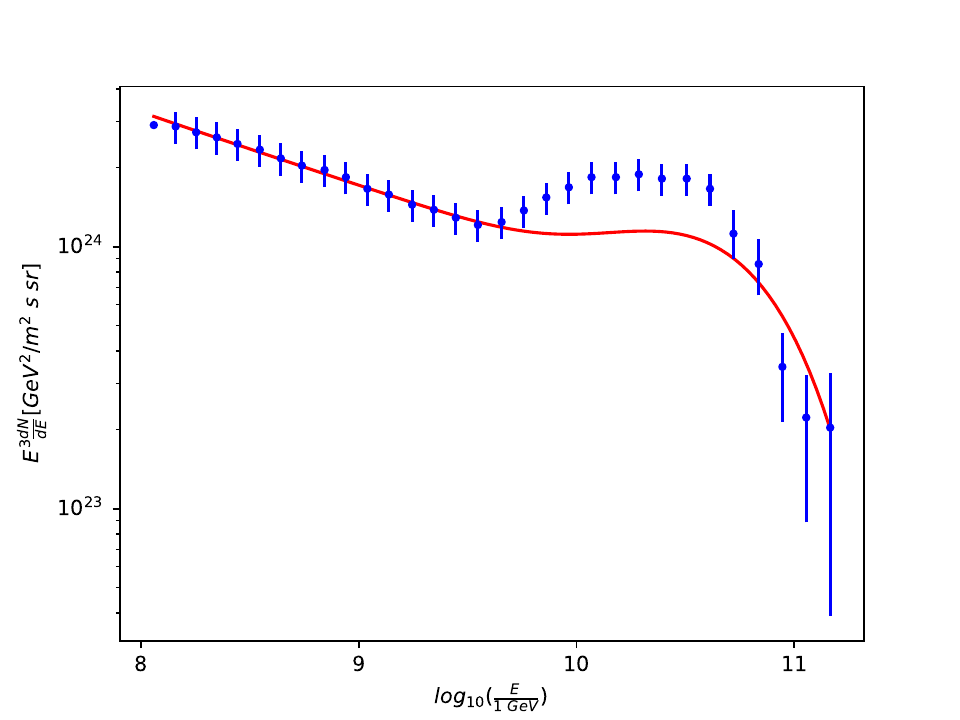}
     \caption{The plot of cosmic rays spectrum in high energy range predicted by our model with the observation data. The top panel shows model fitting and data from Telescope Array with systematic error $20\%$ \cite{Ivanov:2020rqn} where the bottom panel shows model fitting and data from Pierre Auger Observatory with systematic error $14\%$ \cite{PierreAuger:2020qqz,PierreAuger:2021ibw,PierreAuger:2021tmd}.
     \label{fig:PlotALL}}
 \end{figure}

The experimental data chosen in this project are Telescope Array \cite{Ivanov:2020rqn} and Pierre Auger Observatory \cite{PierreAuger:2020qqz,PierreAuger:2021ibw,PierreAuger:2021tmd},
since they are at the tail end of the cosmic ray spectrum ($10^9 - 10^{12}$ GeV) representing the UHECR. First, the chi-square is minimized to determine the value of the four parameters $\{\alpha', m_{\gamma'}, \mathcal{I}_{E}^{(0)}, \mathcal{I}_{G}^{(0)}$\}, where the chi-square function is defined as
\begin{align}
    \chi^2 = \sum_{i=1}^N \frac{ \left(I(s_0^i;\alpha', m_{\gamma'}, \mathcal{I}_{E}^{(0)}, \mathcal{I}_{G}^{(0)})-I_{obs}(s_0^i)\right)^2}{\sigma_i^2}.
\end{align}
The variance, $\sigma_i^2$, is taken to be the systematic error of the observation data ($20\%$ for Telescope Array and $14\%$ for Pierre Auger Observatory). The model flux from the best fit parameters comparing with the experimental data is shown in Fig.(\ref{fig:PlotALL}). However, it turns out that the chi-square value of the best fit of these parameters is not significantly better than the chi-square value of the best fit of the model with only astrophysical fluxes.

\begin{figure}[!ht]
     \centering     \includegraphics[width=0.5\textwidth]{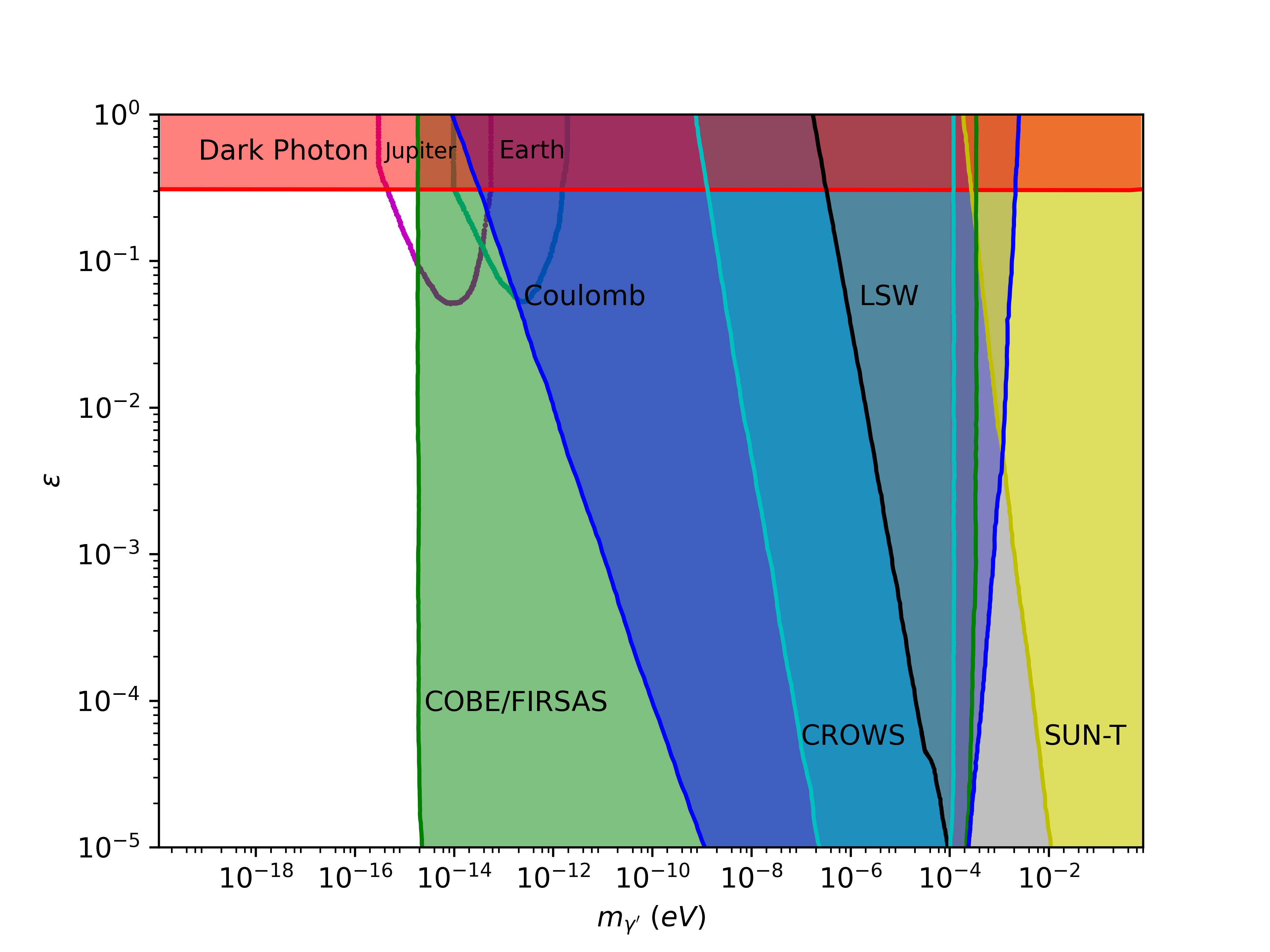}
     \caption{The plot show the constraints of dark photon mass ($m_{\gamma'}$) with the kinetic mixing parameter ($\epsilon$). The red area show the limit from dark photon Bremsstrahlung process. The dash and the solid line represent the limit calculated from Pierre Auger and  Telescope Array respectively. Other areas are the constraints from the experiment, COBE/FIRAS\cite{Garcia:2020qrp,Caputo:2020bdy}, Coulomb\cite{Jaeckel:2010xx}, CROWS\cite{Betz:2013dza}, LSW\cite{Ehret:2010mh}, SUN-T\cite{An:2013yfc} , Jupiter\cite{Davis:1975mn, Jaeckel:2010ni}, Earth\cite{Goldhaber:1971mr}}
     \label{fig:Eps}
 \end{figure} 
 
In order to set the constraint on the model with dark photon radiation, we perform the log-likelihood ratio test between the model with dark radiation and the model without dark photon radiation. The test statistic parameter is given by
\begin{equation}
    \lambda = -2\left(\chi^2(\alpha', m_{\gamma'}, \mathcal{I}_{E}^{(0)}, \mathcal{I}_{G}^{(0)}) - \chi^2(\mathcal{I}_{E}^{(0)}, \mathcal{I}_{G}^{(0)})\right).
\end{equation}
The best values of $\mathcal{I}_{E}^{(0)}$ and $\mathcal{I}_{G}^{(0)}$ are chosen from the minimum of the chi-square of the model with no effect of dark photon radiation. Then the parameter space of $\alpha'$ and $m_{\gamma'}$ is scanned for the test statistic with 0.995 significance level. The mixing parameter is then calculated from the dark photon coupling constant using Eq.(\ref{eq:alphamix}) as
\begin{align}
    \epsilon = \sqrt{\frac{\alpha'}{\alpha+\alpha'}}.
\end{align}
The result is shown in Fig.(\ref{fig:Eps}), with the other experimental constraints on dark photon such as photon mass limits from planetary magnetic field \cite{Davis:1975mn, Jaeckel:2010ni, Goldhaber:1971mr}, limit from CMB spectrum (COBE/FIRAS) \cite{Garcia:2020qrp,Caputo:2020bdy}, limit from deviation of Coulomb's law \cite{Jaeckel:2010xx}, limit from Light Shining through a Wall (LSW) experiment \cite{Ehret:2010mh}, CERN Resonant Weakly Interacting Sub-eV Particle Search (CROW) \cite{Betz:2013dza}, and solar lifetime (SUN-T) \cite{An:2013yfc}. It is worth noting that the effect of dark photon bremstrahlung has been studied in the context of $\pi^0$ decay which gives the constraints that are relevant to a higher mass scale \cite{Blumlein:2013cua}. 
Although the soft dark photon radiation effect alone would prefer a lower dark photon mass and a strong mixing as indicated in the losing fraction, the result from the Telescope Array and Pierre Auger data indicates the strong presence of hard radiation on the UHECRs. This is because the shape of the limit resembles the losing fraction in the hard dark photon Bremsstrahlung case, i.e., the effect of hard radiation increases with the strength of the dark photon coupling, independent of the dark photon mass. The result suggests that the strong mixing parameters $\epsilon > 0.3$ have been ruled out.
The constraint from Telescope Array is slightly stronger than the one from Pierre Auger. This is because the model fluxes fit better with data from Telescope Array. Bremstrahlung covers a wide range of mass and it provides the strongest constraint in the region of ultralight dark photon ($m_{\gamma'} < 10^{-15}\;\;\text{GeV}$). Note that our results are comparable with the constraint derived from the inverse Compton-like scattering of ultralight dark photon \cite{Su:2021jvk}. Recently there are evidences that a significant fraction of cosmic rays at high energy could be nuclei as heavy as iron \cite{PierreAuger:2010ymv}. In this case, the dark photon Bremsstrahlung effect for the heavy nuclei will be enhanced by the factor $Z^2$, where $Z$ is the atomic number of the nuclei. However, we do not consider this effect in this study since the percentage of heavy nuclei in the UHECR is still largely undetermined.



\section{Conclusion}
\label{sec-4}

In this work, we suggest a new method to investigate the potential existence of a dark photon through its kinetic mixing with usual photon. By applying the standard calculation for soft photon radiation, we show that the Bremsstrahlung process for dark photons could result in significant energy loss for protons scattering. This effect can be compared to observational data of ultra-high-energy cosmic rays to test for the presence of dark photons. The UHECR data from Telescope Array and Pierre Auger Observatory data were used to determine the value of four parameters, $\{\alpha', m_{\gamma'}, \mathcal{I}_{E}^{(0)}, \mathcal{I}_{G}^{(0)}\}$, by minimizing the chi-square function. Using the log-likelihood ratio test between the model with dark radiation and the model without dark photon radiation, the constraint from dark photon Bremsstrahlung effect is derived. The result shows that the constraint is strongest in the region of ultralight dark photon.

\appendix

\acknowledgments
We thank Armando di Matteo for helpful suggestions. This work is supported by Thailand NSRF via PMU-B [grant number B05F650021]. CP is also supported by Fundamental Fund 2566 of Khon Kaen University and Research Grant for New Scholar, Office of the Permanent Secretary, Ministry of Higher Education, Science, Research and Innovation under contract no. RGNS 64-043. CP and DS are supported by the National Astronomical Research Institute of Thailand (NARIT).

\bibliographystyle{apsrev4-2}
\bibliography{references.bib}

\end{document}